**What is a tunnel? A comment on frequent imprecision when speaking about tunnels, at the example of two representative structures in the octahedral molecular sieve group.**

Miguel Gregorkiewitz

Department of Physical, Earth and Environmental Sciences, University of Siena, via Laterina 8, I-53100 Siena, Italy.

E-mail: gregormigu@gmail.com.

ORCID: https://orcid.org/0000-0001-8566-4401

**Abstract**

A paper by Su et al (2013)[1] is focused on the presentation of β-$MnO_2$ (pyrolusite) as a new high-capacity electrode for Na batteries, arguing that there is plenty of place in the tunnels to host sodium, which is clearly contrary to general knowledge. Similar imprecisions are found in other work so that it might be useful to shortly address the issue of cavities and tunnels in a general way. Here, two structures of the octahedral molecular sieve group (β-$MnO_2$ pyrolusite and α-$MnO_2$ hollandite) are taken as an example and analysed at some depth regarding their tunnel structure using advanced software. Tunnels and their occupation are of fundamental importance for the application of such materials as electrodes or in strongly correlated electron systems.

"The excellent electrochemical performance of β-MnO$_2$ nanorods could be ascribed to the compact and dense (1 × 1) tunnel-structure in β-MnO$_2$ crystals. The radius of Na ion (1.02 Å) is much smaller than the size of the (1 × 1) tunnel (2.3 Å × 2.3 Å)[30,31]. Therefore, Na ions can facilely insert and extract along the (1 × 1) tunnels in β-MnO$_2$ nanorods."

The preceding string is cited from Su et al. (2013[1], for the references see original paper). Confusion is made with the use of radius and diameter and, more importantly, between interatomic distances and purely geometric measures, including projections, which serve to describe a cavity or bottleneck. In fact, the above cited distances of 2.3 Å × 2.3 Å are normally[2,3] used to give an idea of the dimensions of the 1 × 1 tunnel and refer to the width of the octahedral chain projected on the (001) plane (see Fig. 1c). This width is derived from an O-O distance measured between the centres (nuclei) of two oxygen atoms, so conceptually, there is no way to compare this with ionic radii or diameters used in space-filling considerations.

In the following, I will try to re-assess some terminology and give a detailed description of the tunnel structure in the octahedral molecular sieve frameworks of pyrolusite and hollandite, with the help of advanced software. Materials in this family are presently much studied for their use as insertion electrodes in new batteries and, due to mixed valence Mn$^{3+}$/Mn$^{4+}$ and pronounced structural anisotropy, they are also interesting as strongly correlated electron systems[4]; for both applications, knowledge of the structure and occupation of the tunnels is fundamental.

Octahedral molecular sieves or tunnel oxides are a well known structural family to which many manganese oxides belong. All of these structures are formed by edge-sharing octahedral chains running along **c** (Pasero 2005[5]). In β-MnO$_2$ (pyrolusite, Fig. 1c), single chains share their corners to form 1 × 1 tunnels, and in α-MnO$_2$ (usually cited as hollandite framework), there are double chains forming 2 × 2 tunnels (Fig.s 1a and b). To date, six further (large) tunnel sizes are known in the family (1 × 2, 2 × 3, 2 × 4, 2 × 5, 3 × 3 and 3 × 4). For all structures with large tunnels, many guests have been reported to occupy the cavities but for pyrolusite, cavities in the 1 × 1 tunnels are very small and no stable structures with guests are known with the exception of MnOOH manganite[6] where H occupies a site between two oxygen atoms facing the tunnel at 2.60 Å, and even for the ramsdellite structure (1 × 2 tunnel), only Li is known to occupy the cavity, in a distorted tetrahedron with $d$(Li-O) = 2.01 Å[7].

**What are tunnels?**

A pore system is generally defined by cavities and bottlenecks between adjacent cavities. The cavities are lined with ligands like oxygen which serve to hold the guest atoms somewhere in their interior, related with the stability of the compound which is seen in ion exchange isotherms, BET nitrogen adsorption measures, phase transitions and the like. The bottlenecks, on the other hand, are the windows defining the width of the tunnel along which intracrystalline diffusion of the guest may take place and which is related with an activation energy obtained from diffusion or ionic conductivity experiments. Tunnels may form as a linear 1D sequence of cavities, but there are also zig-zag tunnels and crossing tunnel systems.

To describe a tunnel, we consider its windows and use either topology, such as the ring sizes given above (e.g. 1 × 1 for a ring made up by four corner sharing octahedra), or a metrical measure which gives the dimensions in Ångström.

In hollandite, e.g., the window has topology 2 × 2 and its dimensions are usually[2,3] cited as 4.6 × 4.6 Å², the projected width of the double chain. The cavity corresponding to the 2 × 2 tunnel is a cube shortened along **c** with 3.58 × 3.58 × 2.85 Å³ (cf Fig. 1a), coordinating the centre at 2.90 Å which is optimal for potassium in coordination VIII (2.90 – 1.39 = 1.51 Å, exactly the radius given by Shannon[8]). The window is a square with diagonal 5.06 Å which leaves an open radius of 2.53 – 1.39 = 1.14 Å, a little small for potassium with $r(K^{IV})$ = 1.36 Å, but open for sodium and silver or smaller ions.

The 1 × 1 window in pyrolusite, which is also present in hollandite, is usually cited to have the (projected) width 2.3 × 2.3 Å². The cavity in this tunnel is a tetrahedron flattened along **c** with two edges of 3.33 Å (⊥**c**) and four edges of 2.75 Å (Fig. 1a), and with a radius of 1.81 Å which leaves an open radius of 1.81 – 1.36[1] = 0.45 Å, much smaller than the Shannon[8] radius of sodium, $r(Na^{IV})$ = 0.99 Å, or even lithium, $r(Li^{IV})$ = 0.59 Å. Sodium is therefore never expected to be found in the pyrolusite tunnels: from DFT calculations[9], the 1 × 1 cavity has a 0.276 eV higher energy than the 2 × 2 cavity. For Li, the energy difference is much lower (0.108 eV) and indeed, there is experimental evidence[10-12] that Li might be reversibly inserted in the 1 × 1 cavities under very special circumstances, as well as there exist $Li_xMO_2$ rutile structures based on octahedral M ions with larger radii than Mn ($r$ = 0.60-0.65 instead of 0.53 Å[13,14]).

The bottleneck (window) of the 1 × 1 tunnel is defined by the triangular face of the tetrahedron (3.33, 2.75 and 2.75 Å) which leaves a free radius of 0.37 Å, resulting in a zig-zag tunnel along **c**. Such complicate tunnel structures can be described with appropriate

---

1  Note that the oxygen radius here is for O(III), and for O(V) in the case of the 2 × 2 structure above software[15-20] (for others see reference[20]) which give the exact measures of cavities and bottlenecks along with a graphical representation (e.g. Fig. 2).

Fig. 2 compares the tunnels along **c** in the two cases. The 2 × 2 tunnel is linear and easily described by the cavities and bottlenecks, whereas the small and contorted 1 × 1 tunnel is difficult to visualize and not well separated from alternative tunnels running perpendicular to **c** (there are tetrahedral cavities, evidenced in Fig.s 1b and c, only little smaller with $r$ = 1.67 instead of 1.81 Å, the radius defined by the 1 × 1 tunnels). There is an interesting case of stuffed derivative of the pyrolusite-rutile structure, $Cu_4CaP_2$[23], where the tetrahedra $Cu_4$ fill a much enlarged[2] 1 × 1 tunnel with Cu occupying approximately the pockets seen in Fig. 2. From the picture of the 1 × 1 tunnel in Fig. 2, this compound might have been predicted.

In the preceding, the two octahedral molecular sieve structures of hollandite and pyrolusite were used to demonstrate new ways for an exhaustive description of their tunnel systems, with the help of modern software most of which is freely available. It is hoped that such considerations will become standard in the future, similar to common practice in the tunnel structures of zeolites[24].

**Acknowledgements**

I thank Ángel Herráez and Robert Hanson for their kind help in the use of program *Jmol*.

---

[2] The $CaP_6$ octahedra have a Ca-P bond length of 2.94 instead of 1.91 Å for Mn-O

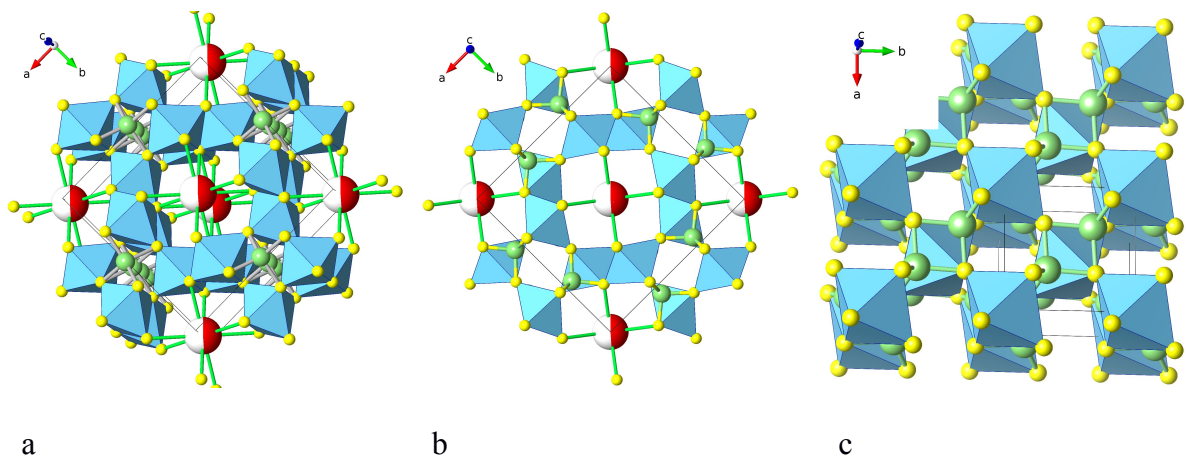

Fig. 1. K-hollandite (a) with 2 × 2 and 1 × 1 tunnels evidenced , and (b) with the 2 × 2 and the small perpendicular tunnels evidenced; manganese octahedra sky blue, K red (half-filled), oxygen yellow, polyhedral bonds green. (c) Pyrolusite with the small perpendicular tunnels evidenced between the 1 × 1 tunnels. There are no cations filling the 1 × 1 tunnels in either hollandite nor pyrolusite, the green spheres just indicate the centres of the respective tunnels. Graphics prepared using VESTA[21].

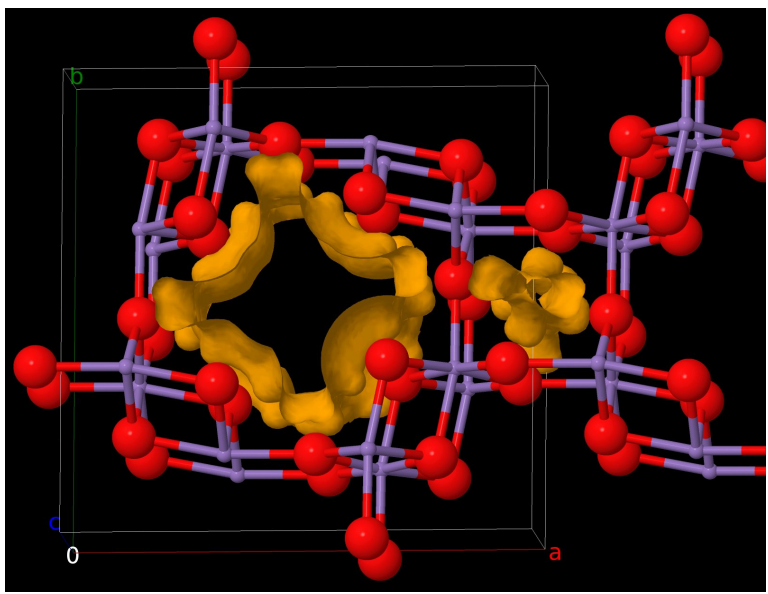

Fig. 2. Representation of tunnels in α-MnO$_2$, showing a 2 × 2 and a 1 × 1 tunnel. The 2 × 2 tunnel is the big black hole in the large ring at left, and the 1 × 1 tunnel is actually reduced to a small, almost invisible, zig-zag tunnel parallel **c** with lateral pockets. Graphics prepared using Jmol[22], as specified in the Appendix (isosurface at 1.37(0.53) + 0.31 Å from O(Mn), respectively).

**Appendix**

For convenience, the Jmol code to produce the cavity image in Fig. 2 is included here.

```
$ load myhollanditeframework.cif {2,1,2}; set window 800 800

$ connect delete; connect 2.2 (_Mn) (_O); connect single radius 0.10

$ {_Mn}.radius=0.18; {_O}.radius=0.46

$ {_Mn}.vdWradius=0.53; {_O}.vdWradius=1.37

$ isosurface resolution 10 pocket cavity 0.31; isosurface subset [2, 5]

$ write IMAGE 2400 2400 JPG 85 "2x2+1x1pocket.31"
```